\documentclass{PoS}
\newcommand{\bea}{\begin{eqnarray}}    
\newcommand{\eea}{\end{eqnarray}}      
\newcommand{\be}{\begin{equation}}
\newcommand{\ee}{\end{equation}}
\newcommand{\bef}{\begin{figue}}
\newcommand{\eef}{\end{figure}}
\newcommand{\cd}{{\langle n(r) \rangle_p}}

\title{Super-homogeneity and inhomogeneities in the large scale matter distribution}

\ShortTitle{Super-homogeneity and  inhomogeneities in matter distribution}

\author{\speaker{Francesco Sylos Labini}\\ 
        Museo Storico della Fisica e Centro Studi e Ricerche Enrico
  Fermi, - Piazzale del Viminale 1, 00184 Rome, Italy \& Istituto dei
  Sistemi Complessi CNR, - Via dei Taurini 19, 00185 Rome, Italy \\
        E-mail: \email{Francesco.SylosLabini@roma1.infn.it}}

\abstract{Super-homogeneity is a property that is supposed to be
  satisfied by matter fluctuations in all standard theoretical models
  of structure formation, such as LCDM and its variants. This is a global
  condition on the correlation properties of the matter density
  field, which can be understood as a consistency contraint in the
  framework of FRW cosmology, and it  corresponds to a very fine
  tuned balance between negative and positive correlations of density
  fluctuations and to the fastest possible decay of the normalized
  mass variance on large scales. By considering several galaxy
  samples, we discuss that these are characterized by the presence of
  large amplitude fluctuations with spatial extension limited only the
   size of the current samples. There is therefore a tension
  between the standard prediction of super-homogeneity and the
  detection of large scale inhomogeneities in the matter distribution
  at scales of the order of 100 Mpc/h.  We discuss the theoretical
  implications of these results with respect to models of structure
  formation and to future galaxy and CMBR data, emphasizing the
  central role of the super-homogeneity property in the current
  description of fluctuations in FRW models.}

\FullConference{International Workshop on Cosmic Structure and
  Evolution - Cosmology2009,\\ September 23-25, 2009\\ Bielefeld,
  Germany}

\begin{document}

\section{Introduction}

More than twenty years ago it has been surprisingly discovered that
galaxy velocity rotation curves remain flat at large distances from
the galaxy center while the density profile of luminous matters
rapidly decays (e.g. \cite{rubin80}).  This is one of the strongest
indications of the need from dynamically dominant dark matter in the
universe. Most attention has been focused on the fact that these bound
gravitational systems contain large quantities of unseen matter and an
intricate paradigm has been developed in which non-baryonic dark
matter plays a central role not only in accounting for the dynamical
mass of galaxies and galaxy clusters \cite{fg79} but also for
providing the initial seeds which have given rise to the formation of
structure via gravitational collapse \cite{pee82}. 

In current standard cosmological models, different forms of dark
matter are needed to explain a number of different phenomena. In fact,
the results of several observations, such as the scale size of
fluctuations of the Cosmic
Microwave Background Radiation (CMBR)
(e.g., \cite{wmap3}), the measurements of clustering mass on large
scales (e.g., \cite{lrg}), the magnitude-redshift relation of
type Ia supernovae (SNe Ia) (e.g., \cite{perl99}), are interpreted to
give consistent measurements of the amount of dark matter.  In this
framework, baryons ($\Omega_B$), which can be detected in the form of,
for example, luminous objects such as stars and galaxies, would only
be the $5\%$ of the total mass in the universe; the rest is made of
entities about which very little is understood: dark matter and dark
energy.  More specifically dark matter, in form of non-baryonic
elementary particles, would contribute to the $\sim 30\%$ of the total
mass of universe ($\Omega_m \sim 0.3$). It is worth noticing that its
direct detection in laboratory experiments is still lacking and that
the standard model of particle physics does not predict the existence
of candidate dark matter particles with the necessary properties from
a cosmological point of view.

Several evidences, from supernovae and other observations, shows that
the expansion of the Universe, rather than slowing because of gravity,
is increasingly rapid. Within the standard cosmological framework,
this must be due to a substance, which has been termed dark energy,
that behaves as if it has negative pressure. This is a mysterious form
of energy which would cause the accelerating expansion of the universe
and it should account for about $70\%$ (i.e.  $\Omega_\Lambda \sim
0.7$) of the mass-energy in the Universe. It is thus not surprising
that great observational and theoretical effort  is
devoted to the understanding of the nature and properties of dark
matter and dark energy which,  giving the main contribution to the
mass-energy density of the universe, play a crucial role, for
example, in the problem of structure formation.

The previous discussion enlightens the fact that we know very little
about the nature of cosmological dark matter both from a fundamental
and observational points of view. Although dark matter is so central
in modern cosmology its amount and properties can only be defined a
posteriori. In this context a crucial question concerns a possible
clear property of dark matter density fields which is not arbitrary,
i.e. a property which has to be satisfied by dark matter fluctuations
under some very general theoretical conditions. In fact, from the
above discussion it seems that much freedom is left for the choice of
dark matter, its physical properties and its statistical
distribution. However there is an important constraint which must be
valid for any kind of initial matter density fluctuation field in the
framework of Friedmann-Robertson-Walker (FRW) models and which
represents a consistency condition to be satisfied by any fluctuation
field compatible with the FRW metric.  As we discuss below this must
be imprinted both in the fluctuations of the CMBR and in the large
scale distribution of galaxies. This is represented by the condition
of super-homogeneity, corresponding in cosmology to the so-called
condition of ``scale-invariance''\footnote{Note that in statistical
  physics the term ``scale invariance'' is used to describe the class
  of distributions which are invariant with respect to scale
  transformations. For instance a magnetic system at the critical
  point of transition between the paramagnetic and ferromagnetic
  phase, shows a two-point correlation function which decays as a
  non-integrable power law. The meaning of ``scale-invariance'' in the
  cosmological context is therefore completely different, referring to
  the property that the mass variance at the horizon scale be constant
  (see below).}\cite{glass}.

\section{Super-homogeneity in LCDM} 
According to standard theoretical models derived from inflationary
mechanisms, the most prominent feature of the matter density field in
the early universe is that it presents super-homogeneous features on
large enough scales~\cite{glass}.  To clarify the meaning of this
condition, let us consider the properties of statistically homogeneous
and isotropic stochastic processes, describing the matter density
field and its fluctuations. Let us firstly start with the simplest
stochastic point process: the Poisson distribution.  In this case,
particles are placed completely randomly in space (i.e. without
correlations), and mass fluctuations in a sphere of radius $R$ growth
as $R^3$, i.e. like the volume of the sphere. This is thus a uniform
(i.e. spatially homogeneous), statistically homogeneous and isotropic
(i.e. stationary) distribution.  In addition to these properties, a
super-homogeneous distribution shows the peculiar feature that mass
fluctuations grow in the slowest possible way, i.e. slower than $R^3$
\cite{glass,book}.  To be more precise, let us introduce the
normalized mass variance
\begin{equation}     
\sigma^2(R)=      
\frac{\langle M(R)^2 \rangle - \langle M(R) \rangle^2}   
{\langle M(R) \rangle^2}\,,     
\label{v1}     
\end{equation}  
where $\langle M(R) \rangle$ is the average mass in a {\it sphere} of
radius $R$ and $\langle M(R)^2 \rangle$ is the average of the square
mass in the same volume\footnote{Hereafter we consider only the case
  in which the variance is computed in a sphere of radius
  $R$. Sometimes in the literature Gaussian spheres are used; while
  this choice does allow a mathematically coherent formulation, from a
  physical point of view  it hides the important properties of
  super-homogeneous distributions (see discussion in \cite{glass}).}.
Given that for uniform systems $\langle M(R) \rangle \sim R^3$, for a
Poisson distribution we find
\be
\label{vpoi}
\sigma^2(R) \sim R^{-3} \;.  
\ee 
On the other hand, for
super-homogeneous systems the variance behaves as 
\be
\label{vord}
\sigma^2(R) \sim R^{-4} \;, 
\ee 
which is the fastest possible decay
for discrete or continuous distributions \cite{glass}. Thus, the
super-homogeneous nature of matter distribution corresponds to the
presence of mass fluctuations which are depressed with respect to the
uncorrelated Poisson case.

For example a perfect cubic lattice of particle is a super-homogeneous
system, although this is not a stationary stochastic point process
because of its intrinsic symmetries.  In the former class, for
instance, we find the one component plasma (OCP), a well-known system
in statistical physics~\cite{lebo}.  The OCP is simply a system of
charged point particles interacting through a repulsive $1/r$
potential, in a uniform background which gives overall charge
neutrality. At thermal equilibrium, and for high enough temperatures,
the spatial configuration of charged particle is super-homogeneous
(i.e. the glassy configuration).  Simple modifications of the OCP can
produce equilibrium correlations of the kind assumed in the
cosmological context, as for instance in the LCDM model \cite{lebo}.

 In the cosmological context the super-homogeneous nature of matter
 density fluctuations in the early universe, was firstly hypothesized
 in the seventies \cite{harrison,zeldovich}.  It subsequently gained
 in importance with the advent of inflationary models in the eighties,
 and the demonstration that such models quite generically predict a
 spectrum of fluctuations of this type. 
The reason for this
 peculiar behavior of primordial density fluctuations is the
 following.  In a FRW cosmology there is a fundamental characteristic
 length scale, the horizon scale $R_H(t)$. It is simply the distance
 light can travel from the Big Bang singularity $t=0$ until any given
 time $t$ in the evolution of the Universe, and it grows linearly with
 time.  Harrison \cite{harrison} and Zeldovich \cite{zeldovich}
 introduced the criterion that matter fluctuations have to satisfy on
 large enough scales. This is named the Harrison-Zeldovich criterion
 (H-Z), and it can be written as
\be \sigma_M^2 (R=R_H(t)) = {\rm constant}.
\label{H-Z-criterion}
\ee
This conditions states that the mass variance at the horizon scale is
constant: this can be expressed more conveniently in terms of the
power spectrum (PS) of density fluctuations
\cite{glass}
\be
P(\vec{k})=\left<|\delta_\rho(\vec{k})|^2\right>
\label{ps-defn}
\ee where $\delta_\rho(\vec{k})$ is the Fourier Transform of the
normalized fluctuation field $(\rho(\vec{r})-\rho_0)/\rho_0$, being
$\rho_0$ the average density. It is possible to show that
Eq.\ref{H-Z-criterion} is equivalent to assume $P(k) \sim k$ (the H-Z
PS).  In particular the initial fluctuations are taken to have Gaussian
statistics and a spectrum which is exactly, or very close to, the
so-called H-Z PS; in this situation matter distribution present
fluctuations of the type given by Eq.\ref{vord} \cite{glass}. Since
the fluctuations are Gaussian, the knowledge of the PS gives a
complete statistical description of the fluctuation field.

Let us briefly frame super-homogeneous systems comparing them to the
different uniform and stationary distributions.  Without loss of
generality, let us suppose that $P(k)=Ak^nf(k)$, where $A>0$ and
$f(k)$ a cut-off function chosen such that (i) $\lim_{k\rightarrow 0}
f(k) = 1$, and (ii) $\lim_{k\rightarrow \infty} k^n f(k)$ is finite.
We also require $n>-3$ to have the integrability of $P(k)$ around $k=
0$ \cite{book}.  It is then possible to proceed to the following
classification for the scaling behavior of the normalized
mass-variance in real space spheres \cite{glass,book}:
\begin{equation}
\sigma^2(R)\sim\left\{
\begin{array}{ll}
R^{-(3+n)} & \mbox{for }\, -3<n<1\\
R^{-(3+1)}\log R & \mbox{for }\, n=1\\
R^{-(3+1)} & \mbox{for }\, n>1
\end{array}
\right. \;. 
\label{sigma5}
\end{equation}
For $-3<n<0$ (i.e., $P(0) = +\infty$), we have ``super-Poisson'' mass
fluctuations typical of systems at the critical point of a second
order phase transition~\cite{book}.  For $n=0$ (i.e., $P(0) = A >0$),
we have Poisson-like fluctuations, and the system can be called {\em
  substantially Poisson}.  This behavior is typical of many common
physical systems e.g., a homogeneous gas at thermodynamic equilibrium
at sufficiently high temperature. Finally for $n\ge 1$ (i.e., $P(0)=0$),
we have ``sub-Poisson'' fluctuations, and thus {\em super-homogeneous}
systems \cite{glass,book}.

In order to illustrate more clearly the physical implications of the
H-Z condition, one may consider gravitational potential fluctuations
$\delta\phi(\vec{r})$ which are linked to the density fluctuations
$\delta\rho(\vec{r})$ via the gravitational Poisson equation:
%
%\begin{equation} 
$\nabla^2\delta\phi(\vec{r})=4\pi G \delta\rho(\vec{r})\,. $
%\label{poii} 
%\end{equation} 
%
From this, transformed to Fourier space, it follows that the PS of the
potential $P_{\phi}(k)=\left<|\delta\hat\phi(\vec{k})|^2\right>$ is
related to the density PS $P(k)$ through the equation
\be
P_{\phi}(k)\sim \frac{P(k)}{k^4}\,.
\ee
The H-Z condition corresponds therefore to $P_{\phi}(k) \propto
k^{-3}$. In this case, the variance of the gravitational potential
fluctuations is $\sigma_{\phi}^2(R) \approx \frac{1}{2} P_{\phi}(k)
k^3|_{k =R^{-1}} $ \cite{glass}.
The H-Z condition fixes this variance to be constant as a function of
$R$.  This is a {\it consistency constraint} in the framework of FRW
cosmology. Indeed, the FRW is a cosmological solution for a perfectly
homogeneous Universe, about which fluctuations represent an
inhomogeneous perturbation. If density fluctuations obey to a
different condition than Eq.\ref{H-Z-criterion}, then the FRW
description will always break down in the past or future, as the
amplitude of the perturbations become arbitrarily large or small.  For
this reason the super-homogeneous nature of primordial density field
is a fundamental property independently on the nature of dark
matter. This is a very strong condition to impose, and it excludes
even Poisson processes ($P(k)=$const. for small $k$) \cite{glass}:
indeed, in this case the fluctuations in the gravitational potential
may diverge at large scales.

Various models of primordial density fields differ for the behavior of
the PS at large wave-lengths, i.e. at relatively small scales,
depending on the specific properties hypothesized for the dark matter
component. For example, for the case the Cold Dark Matter scenario
(CDM), where elementary non-baryonic dark matter particles have a
small velocity dispersion, the PS decays as a power law $P(k) \sim
k^{-2}$ at large $k$. For Hot Dark Matter (HDM) models, where the
velocity dispersion is large, the PS presents an exponential decay at
large $k$. However at small $k$ they both exhibit the H-Z tail $P(k)
\sim k$ which is indeed the common feature of all density fluctuations
compatible with FRW models. The scale $r_c \approx k_c^{-1}$ at which
the PS shows the turnover from the linear to the decaying behavior is
fixed to be the size of the horizon at the time of equality between
matter and radiation \cite{pee82}.

In terms of correlation function $\xi(r)$ (the Fourier conjugate of
the PS) CDM/HDM models present the following behavior for the early
universe density field. This is positive at small scales, it crosses
zero at a certain scale and then it is negative approaching zero with
a tail which goes as $-r^{-4}$ (in the region corresponding to $P(k)
\sim k$) \cite{book}. The super-homogeneity (or H-Z) condition
corresponds to the following limit condition 
\be
\label{intcos} 
\int_0^{\infty} d^3r \xi(r) = 0 \;,
\ee
which is another way to reformulate the condition that $\lim_{k
  \rightarrow 0} P(k) = 0$, i.e.  $P(0)=0$.  This means that there is
a fine tuned balance between small-scale positive correlations and
large-scale negative anti-correlations~\cite{glass,book}.
This is the behavior that one would like to detect in the data in
order to confirm inflationary models. Note that the Eq.\ref{intcos} is
different, and much stronger, from the requirement that any {\it
  uniform} stochastic process has to satisfy, i.e. $\lim_{R
  \rightarrow \infty} \sigma^2(R) = 0$
\cite{book}.

It is worth noticing that the physical meaning of the constraint
$P(0)=0$ is often missed in the cosmological literature because of a
confusion with the so-called ``integral constraint'' , which is
another apparently similar, but actually completely different
constraint.  This latter constraint holds for the {\it estimator} of
the two-point correlation function in a finite sample, and it may take
a form similar to the condition $P(0)=0$ defining super-homogeneous
distributions, but over a {\it finite} integration volume. These two
kinds of constraint have a completely different origin and meaning,
one ($P(0)=0$) describing an intrinsic property of the fluctuation
field in a well-defined class of distributions, the other a property
of the estimated correlation function of any distribution as measured
in a finite sample.  Their formal resemblance however is not
completely without meaning and can be understood as follows: in a
super-homogeneous distribution the fluctuations between samples are
extremely suppressed, being smaller than Poisson fluctuations; in a
finite sample a similar behavior is artificially imposed since one
suppresses fluctuations at the scale of the sample by construction by
measuring fluctuations only with respect to the estimation of the
sample density (see discussion below) \cite{glass,book}.

The super-homogeneity prediction is fixed in the early universe
density field which should be represented by CMBR anisotropies. There
are two additional physical elements which must be considered for what
concerns the matter density field we observe today in the form of
galaxies: (i) evolution due to gravitational clustering and (ii)
biasing \cite{durrer,cdmtheo}. Let us briefly discuss these two
issues.

(i) Fluctuations in the matter density field provide the source of the
Poisson equation for the formation of structures. In LCDM models, this
occurs in a bottom-up manner, i.e. structures at small scales are
formed first and then larger and larger scales collapse.  In the
linear regime it is possible to work out the solution to the
Vlasov-Poisson system of equations in an expanding universe
\cite{pee82}. In this case it is easily found that fluctuations are
linearly amplified during the linear phase of gravitational collapse.
Given the extremely fine tuning of correlations characterizing a
super-homogeneous distribution one may wonder whether the growth of
small scales non-linear structures may introduce some distortions of
the PS at large scales.  An argument, firstly discussed by Zeldovich
\cite{zeldo} and recently refined by \cite{viot}, states that the
perturbations to a mass distribution introduced by moving matter
around on a finite scale $r_f$, while preserving locally the center of
mass and momentum, lead to a modification to the PS at small $k$
(i.e. smaller than the inverse of the characteristic length scale
$r_f^{-1}$) which is proportional to $k^4$. Since, as we have seen
above, the matter distribution has a PS which is proportional to $k$
at small $k$, this is not distorted by non-linearities at small
scales.  The scale of non-linearity in current models is placed at
$\sim 10$ Mpc/h, and on larger scales the correlation function is only
linearly amplified with respect to that of the initial conditions. For
this reason, gravitational clustering does not break the
super-homogeneous nature of matter distribution.

(ii) In standard models of structure formation galaxies result from a
{\it sampling} of the underlying CDM density field: for instance one
selects (observationally) only the highest fluctuations of the field
which would represent the locations where galaxy will eventually
form. It has been shown that sampling a super-homogeneous fluctuation
field changes the nature of correlations~\cite{durrer,cdmtheo}. The reason can
be found in the property of super-homogeneity of such a distribution:
the sampling necessarily destroys the surface nature of the
fluctuations, as it introduces a volume (Poisson-like) term in the
mass fluctuations, giving rise to a Poisson-like PS on large scales
$P(k)\sim$ constant.  The ``primordial'' form of the PS is thus not
apparent in that which one would expect to measure from objects
selected in this way. This conclusion should hold for any generic
model of bias and its quantitative importance has to established in
any given model~\cite{durrer}.  On the other hand one may
show~\cite{durrer,cdmtheo} that the negative $r^{-4}$ tail in the correlation
function does not change under sampling: on large enough scales, where
in these models (anti) correlations are small enough, the biased
fluctuation field has a correlation function which is linearly
amplified with respect to the underlying dark matter correlation
function. For this reason the detection of such a negative tail would
be the main confirmation of the super-homogeneous character of
primordial density field ~\cite{book}.

To conclude this brief summary about the statistical
properties of  standard models, we mention the
baryon acoustic oscillations. The physical description which gives
rise to these oscillations is based on fluid mechanics and gravity:
when the temperature of the CMBR was  hotter than $\sim 1000$ K, photons
were hot enough to ionize hydrogen so that baryons and photons can be
described as a single fluid.  Gravity attracts and compresses this
fluid into the potential wells associated with the local density
fluctuations. Photon pressure resists this compression and sets up
acoustic oscillations in the fluid. Regions that have reached maximal
compression by recombination become hotter and hence are now visible
as local positive anisotropies in the CMBR.  The principal point to
note is that while $k-$oscillations are de-localized, in real space
the correlation function shows a characteristic corresponding feature
at a certain well-defined scale.  In particular $\xi(r)$ has a
localized ``bump'' at the scale corresponding to the frequency of
oscillations in $k$ space.  This is not really surprising: it simply
reflects that the Fourier Transform of a regularly oscillating
function is a localized function. Formally the bump of $\xi(r)$
corresponds to a scale where the first derivative of the correlation
function is not continuous \cite{book}.

\section{Galaxy distribution: from inhomogeneity to super-homogeneity ? }

The main information about the matter distribution in the present
universe is derived from the analysis of the correlation properties of
galaxy structures. As mentioned above, in standard models and in the
absence of observational selection effects, Eq.\ref{intcos} should be
satisfied.  However the situation is not so simple and can be
summarized as follows.  On small scales $r < 10$ Mpc/h strong
clustering, driven by the non-linear phase of gravitational dynamics,
should have erased the trace of the initial (linear) matter density
field. On larger scales density fluctuations have only been amplified
by linear gravitational clustering. Thus for $10 < r < 150$ Mpc/h the
correlation function should be positive, crossing zero at about $150$
Mpc/h (the size of the Hubble horizon at the time of equality between
matter and radiation) and then being negative, with a negative power
law tail of the type $\xi(r) \sim - r^{-4}$ at larger scales
\cite{glass,cdmtheo,bao}. In the regime of strong clustering, i.e.  $r
< 10$ Mpc/h, one expects deviation from Gaussian behavior, while at
larger scales the initial Gaussian probability density function of
density fluctuations should be persevered by linear gravitational
clustering.

\subsection{Galaxy correlations: some contradictory results} 

There are several observations pointing toward the fact that galaxy
structures are strongly inhomogeneous at very large scales.  However
there are also measurements which indicate that on large enough scales
fluctuations in the galaxy density field are small.  It seems there is 
a contradictory situation where different authors, employing different
statistical techniques, measure different properties. To sort out 
the reasons behind this we should consider how these measurements 
have been performed. 

There are two different statistical methods to measure fluctuations:
those which determine field-to-field fluctuations or fluctuations as a
function of redshifts, and those which instead measure the amplitude
of {\it relative} fluctuations, i.e. by normalizing the observed
amplitude of fluctuations to the estimation of the sample density. We
discuss some recent results obtained with both methods, emphasizing
the contradictory results which have been obtained by different
authors.  Then in the next section we discuss that this paradoxical
situation can be understood by a careful examination of the
assumptions which enter in both determinations. An analysis of
finite-size effects will ultimately solve this
contradiction \footnote{The interested reader can read
  \cite{2df_epl,2df_aea,sdss_epl,sdss_aea,bao,tibor} for further
  details.}.

The counting of the number of galaxies, in samples with the same
selection effects, is certainly a good although qualitative way to
determine galaxy fluctuations. For instance, recently, there have been
found several evidences of large scale fluctuations (e.g. the
so-called ``local hole'') when counting galaxies as a function of
apparent magnitude in the 2 degree Field Galaxy Redshift Survey and in
the Two Micron All Sky Survey \cite{busswell03,frith03,frith06}: these
show the existence of large scale fluctuations of $30\%$ with a
linear size across the sky of $\sim 200$ Mpc/h.  Similar large scale
fluctuations, extending over several hundreds Mpc have been found in
Sloan Digital Sky Survey. In particular, it has been found that the
apparent number density of bright galaxies increases by a factor
$\approx $ 3 as redshift increases from $z = 0$ to $z = 0.3$
\cite{loveday}. This is again the signature of a coherent change in
the galaxy density field over an enormous range of scale. Whether
galaxy evolution can also be responsible of such a behavior is a
question which must be investigated carefully, as in this case one is
comparing estimation of the local galaxy density as a function of
redshift \cite{sdss_aea}. 

On the other hand, most of standard measurements of galaxy
correlations and fluctuations are based on the calculation of the two
point correlation function $\xi(r)$. For instance in a sample of
luminous red galaxy (LRG) of the Sloan Digital Sky Survey (SDSS) it
was found that fluctuations are of order $10^{-2}$ on scales of $\sim
100$ Mpc/h allowing a determination of the baryonic acoustic peak
followed by the zero-crossing scale of $\xi(r)$ \cite{lrg}.  However
in other samples the situation is even different. For instance in the
2dFGRS it was measured that the zero-crossing scale occurs at $50 $
Mpc/h \cite{martinez}, being thus fluctuations even smaller on larger
scales.

In summary the measurements of galaxy fluctuations seem to show
different and contradictory results when different methods are
used. However even the different behavior of $\xi(r)$ in different
samples should be explained. This can be achieved
through the consideration of finite-size effects. We will give 
a brief introduction to the problem in the next two sections.

\subsection{Large scale fluctuations, large scale inhomogeneity}

An important assumption commonly used in the estimation of the
amplitude and the spatial extension of galaxy correlations is that the
sample average gives a reliable determination of the ``real'' average
density. The determination of the correlation function $\xi(r)$
implies indeed such a normalization.  On very general grounds, this is
a very strong assumption which is not (exactly) satisfied in any
sample. Let us briefly explain why.  The determination of correlation
properties of a given stochastic point process depends on the
underlying correlations of the point distribution itself \cite{book}.
There can be different situations for the statistical properties of
any set of points (in the present case, galaxies) in a finite sample.
Let us briefly consider four different cases
\cite{sdss_aea}.
{\it Inside a given sample} galaxy distribution is well-approximated
by a {\it uniform} stochastic point process, or in other words, {\it
  inside a given sample} the average density is well-defined, i.e. it
gives a reliable estimation of the ``true'' average density (modulo
fluctuations). This means that the density, measured for instance in a
sphere of radius $r$ randomly placed inside the sample, has small
fluctuations. In this situation the relative fluctuations between the
average density estimator and the ``true'' density is smaller than
unity. Density fluctuations maybe correlated, and the correlation
function can be {\it (i)} short-ranged (e.g., exponential decay) or
{\it (ii)} long-ranged (e.g., power-law). In other words these two
cases correspond to a uniform stochastic point process with (i)
short-range and (ii) long-range correlations.

On the other hand it may happen that, inside a given sample, galaxy
distribution is not uniform. In this situation, the density measured
for instance in a sphere of radius $r$ randomly placed inside the
sample, has large fluctuations, i.e. { it wildly varies in different
  regions of the sample.}  In this situation the point distribution
can generally present long-range correlations of {\it large amplitude}
and the estimation of the (conditional \footnote{ Conditional
  statistics are not normalized to the sample density estimation
  (which is a global quantity in a given sample) while they measure
  local statistical properties.}) average density presents a {\it
  systematic} dependence on the sample size.  Then it may present,
case {\it (iii)}, or not, case {\it (iv)}, self-averaging properties
\cite{sdss_aea}, { depending on whether or not measurements of the
  density in different sub-regions show systematic (i.e., not
  statistical) differences} that depend, for instance, on the spatial
positions of the specific sub-regions.  When this is so, the
considered statistics are not statistically self-averaging in
space. In this case, for instance, the probability density function of
fluctuations systematically differs in different sub-regions and
whole-sample average values are not meaningful descriptors. In general
such systematic differences may be related to two different
possibilities: (i) that the underlying distribution is not
translationally and/or rotationally invariant; (ii) that the volumes
considered are not large enough for fluctuations to be self-averaging.
One may perform specific statistical tests to distinguish between
these two possibilities \cite{sdss_aea}.

Concerning the determination of statistical properties, a fundamental
assumption is very often used in the finite-sample analysis: that
the sample density is supposed to provide a reliable estimate of the
``true'' space density, i.e. that the point distribution is
well-represented by the case (i) or (ii) above.  In this situation the
relative fluctuations between the average density estimator and the
``true'' density is smaller than unity.  In general, this is a very
strong assumption which may lead to underestimate finite size effects
in the statistical analysis.  For instance, let us suppose that the
distribution {\it inside the given sample} is {\it not} uniform,
i.e. case (iii) and (iv) above.  In this case the results of the
statistical analysis are biased by important finite-size effects, so
that all estimations of statistical quantities based on the uniformity
assumption (i.e. the two-point correlation function and all quantities
normalized to the sample average) are affected, on all scales, by this
{\it a-priori} assumption which is inconsistent with the data
properties \cite{book}.  In addition, while for the case (iii) one may
consider a class of whole sample averaged quantities, i.e. conditional
statistics, in the case (iv) these become meaningless.

In a series of papers \cite{2df_epl,2df_aea,sdss_epl,sdss_aea,bao} it
was actually found that in the SDSS samples the probability density
function (PDF) of conditional fluctuations (i.e. not normalized to the
estimation of the sample density), filtered on large enough spatial
scales (i.e., $r>30$ Mpc/h), shows relevant systematic variations in
different sub-volumes of the survey. Instead for scales $r<30$ Mpc/h
the PDF is statistically stable, and its first moment presents scaling
behavior with a negative exponent around one. Thus while up to 30
Mpc/h galaxy structures have well-defined power-law correlations, on
larger scales it is not possible to consider whole sample average
quantities as meaningful and useful statistical descriptors. This
situation is due to the fact that galaxy structures correspond to
density fluctuations which are too large in amplitude and too extended
in space to be self-averaging on such large scales inside the sample
volumes: galaxy distribution is inhomogeneous up to the largest
scales, i.e. up to $r \approx 100$ Mpc/h, probed by the SDSS
samples. A similar results was obtained for the 2dFGRS samples. In
addition in \cite{tibor} we showed that in the newest SDSS samples, on
very a large range of scales up to $r \sim 80$ Mpc/h (where
fluctuations in this sample show self-averaging properties), both the
average conditional density and its variance show a nontrivial scaling
behavior, which resembles to criticality. The density depends, for $10
< r < 80$ Mpc/h, only weakly (logarithmically) on the system
size. Correspondingly, we find that the density fluctuations follow
the Gumbel distribution of extreme value statistics. This distribution
is clearly distinguishable from a Gaussian distribution, which would
arise for a homogeneous spatial galaxy configuration. The comparison
between determination of the PDF of conditional fluctuations in
samples of different volumes clearly show the importance of
finite-size effects.

These results are in agreement with the determination of
field-to-field fluctuations and of the redshift distributions. However
they seem to be in contradiction with the measurements of $\xi(r)$:
they are so only in the sense that these determinations are strongly
biased by finite size effects because on the a-priori assumptions on
which they are based, and thus do not allow one to properly measure
fluctuations and correlations of galaxies in the current samples.

%%%%%%%%%%%%%%%%%%

\subsection{Super-homogeneity in the  matter distribution ?}

In order to illustrate the problems related to estimations of the
(possible) super-homogeneous property in future galaxy surveys, let us
briefly discuss some finite-size effects that would affect 
the measurements of the correlation function even in the case
where the sample density is constant as a function of the sample size,
i.e. it does not show a systematic dependence as for the real data.
In this case the sample density differs from the ``real'' average density
(infinite volume limit) because there are finite-size fluctuations.

Let us call $\overline {X(V)}$ the statistical estimator of an average
quantity $\langle X \rangle $ in a volume $V$ (where $\langle X
\rangle $ denotes the ensemble average and $\overline {X}$ the sample
average). To be a valid estimator $\overline {X(V)}$ must satisfy
\cite{book}
 \be
\label{bias1}
\lim_{V\rightarrow \infty} \overline{X(V)}  = \langle X \rangle \;. 
\ee
A stronger condition is that the ensemble average of the estimator, in
a finite volume $V$, is equal to the ensemble average $\langle X
\rangle$:
\be
\label{bias2} 
\langle \overline {X(V)} \rangle = \langle X \rangle \;. 
\ee
An estimator is called unbiased if this condition is satisfied;
otherwise, there is a systematic bias in the finite volume relative to
the ensemble average.  Any estimator $\overline{ \xi(r)}$ of the
correlation function $\xi(r)$, is generally biased. This is because
the estimation of the sample mean density is biased when correlations
extend over the sample size and beyond. In fact, the most common
estimator of the average density is \be
\label{sd} 
\overline n = \frac{N}{V} \;, 
\ee
where $N$ is the number of points in a sample of volume $V$.
It is simple to show that \cite{book} 
\be
\label{biasave}
\langle \overline n \rangle = \langle n \rangle \left( 1 +
\frac{1}{V} \int_V d^3 r \xi(r) \right) \;. 
\ee
Therefore only in case when $\xi(r) =0$ (i.e. for a Poisson
distribution), Eq.\ref{sd} is an unbiased estimator of the ensemble
average density. 

The correlation function can be written, without loss of generality,
as \be \xi(r) \equiv \frac{\langle n(r)n(0)\rangle} {n_0^2} -1 \equiv
\frac{\cd}{n_0}-1 \;, \ee where the conditional density $\cd = \langle
n(r)n(0)\rangle/n_0$ gives the average number of points in a shell of
radius $r$ and thickness $dr$ from an occupied point of the
distribution. Thus the estimator of $\xi(r)$ can be simply written as
\cite{book} \be
\label{xifs1}
\overline{ \xi(r)}  = \frac{\overline{ (n(r))_p}}{\overline n} -1 \,, 
\ee
where $\overline n$ is the estimated number density in the sample and
$\overline{ (n(r))_p}$ is the estimator of the conditional
density. The latter can be written as
\be
\label{cond}
\overline{ (n(r))_p}= 
\frac{1}{N_c(r)} \sum_{i=1}^{N_c(r)} \frac{\Delta N_i (r, \Delta r) }{\Delta V} \;, 
\ee
where $\Delta N_i(r, \Delta r)$ is the number of points in the shell of radius $r$,
thickness $\Delta r$, and volume $\Delta V = 4 \pi r^2 \Delta r$
centered on the $i^{th}$ point of the distribution. Note that the
number of points $N_c(r)$ contributing to the average in Eq.\ref{cond}
is scale-dependent, as   only those points are considered such that
when chosen as a center of the sphere of radius $r$, this is fully
included in the sample volume \cite{cdmtheo} 
\
The sample density can be estimated in various ways. Suppose that the
sample geometry is simply a sphere of radius $R_s$. The most
convenient estimation in this context is to choose
\be
\label{ne1}
\overline n= \frac{3}{4\pi R_s^3} \int_0^{R_s}  \overline{ (n(r))_p} 4\pi r^2 dr \;,
\ee
as in this case the following integral constraint is satisfied 
\be
\label{xifs}
\int_0^{R_s} \overline{ \xi(r)} r^2 dr = 0 \;.  \ee

\begin{figure}
\includegraphics[width=.6\textwidth]{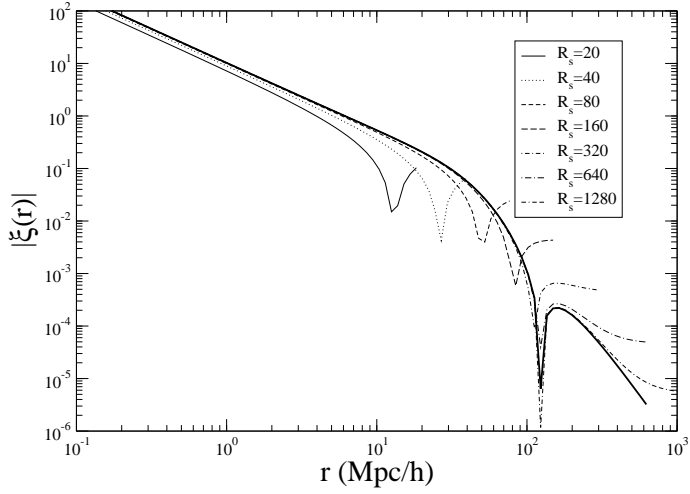} 
\caption{ Expected estimation of the LCDM correlation function in
  samples of different sizes (from \cite{cdmtheo})}
\label{fig1}
\end{figure}

In Fig.\ref{fig1} we show the finite-size effect of the integral
constraint, in samples of different sizes, for the case of a LCDM
correlation function. One may note that that when the sample size is
$R_s <r_c$ (where $r_c$ is the zero-crossing scale) both the amplitude
and the zero-crossing scale are affected by a strong bias. Instead
when $R_s >r_c$ the {\it tail} of the correlation function is
distorted with respect to the ``true'' shape. 

Note that the condition given by Eq.\ref{xifs} is {\it satisfied
  independently of the functional shape of the underlying correlation
  function $\xi(r)$ and for all $R_s$ !}  In addition, note that this
condition holds in a finite sample, while the super-homogeneity
condition (Eq.\ref{intcos}) holds in the {\it infinite} volume
limit. Therefore, in the case in which the difference between the
sample average and the infinite volume limit average is due to
fluctuations (cosmic variance), in order to detect the zero point
properly one must check that this is stable as a function of the
sample size $R_s$.  Another way to look at the standard determinations
of the correlation function previously mentioned, is indeed to check
that the zero point does not change in different samples of different
size. This is in fact the case, and thus our conclusion is that the
measured shape and amplitude of the correlation function is strongly
biased by (uncontrolled) finite size effects.

\section{Super-homogeneity in the Cosmic Microwave Background ?}

Primordial density fluctuations have imprinted themselves not only in
the matter distribution, but also on the patterns of radiation, and
those variations should be detectable in the CMBR.  Three decades of
observations have revealed fluctuations in the CMBR of amplitude of
order $10^{-5}$ \cite{wmap3}.  It is in fact to make these
measurements compatible with observed structures that it is necessary
to introduce non-baryonic dark matter which interact with photons only
gravitationally, and thus in a much weaker manner than ordinary
baryonic matter.  Thus in standard models of structure formation dark
matter plays the dominant role of providing density fluctuation seeds
which, from the one hand are compatible with observations of the CMBR
and from the other hand they are large enough to allow the formation,
through a complex non linear dynamics, of the galaxy structures we
observe today.  In standard cosmological theories the CMBR represents
a bridge between the very early universe and the universe as we
observe today and in particular the galaxy structures. On the one hand
the CMBR probes the very early hot universe at extreme energies
through the theories proposed --- notably ``inflation'' --- to explain
the origin of these perturbations. On the other hand the anisotropies
reflect the local very small amplitude perturbations which give the
initial conditions for the gravitational dynamics which should
subsequently generate the galaxy structures observed today.

In the CMBR one measures fluctuations in temperature on the sky i.e.,
on the celestial sphere. We will not enter here into the detail of the
physical theory in standard models which link these temperature
fluctuations to the mass density field
\cite{padm}.    
It is useful however for what follows to give the precise relation between
the two quantities. 
The temperature fluctuation field
$ \frac{ T(\theta,\phi) - \langle T \rangle }
{ \langle T \rangle } \equiv
\frac{\delta T}{T} (\theta, \phi) \; , $
where $ \theta\;,\phi$ are the two angular coordinates,  is conventionally 
decomposed in spherical harmonics on the sphere:
\be 
\label{cmbr-def} 
\frac{\delta T}{T} (\theta, \phi)= \sum_{l=0}^{\infty} \sum_{m=-\ell}^{+\ell}
a_{\ell m} Y_{\ell m} (\theta, \phi) \;.
\ee 
The variance of these coefficients $a_{\ell m}$ is then 
related to the matter power spectrum through
\index{spherical harmonics}
\be 
\label{cmbr1} 
C_{\ell} \equiv \langle |a_{\ell m}|^2 \rangle = \frac{H_0^4}{2\pi}
\int_0^{\infty} dk \frac{P(k)}{k^2} |j_{\ell}(k\eta)|^2 
\ee 
where $j_\ell$
is the spherical Bessel function and $\eta \simeq 2H_0^{-1}$ is a
constant at fixed time ($H_0$ is the Hubble constant today). Note that
the ensemble average contains no dependence on $m$ because of the
assumption of statistical isotropy.  Taking $P(k) \sim k$ in
(\ref{cmbr1}) we get that the $\ell>2$ multi-poles are given by
%\be 
%\label{cobe6} 
$C_\ell \sim (\ell(\ell+1))^{-1} \;, $
%\ee 
so that the H-Z condition for the power spectrum  $n=1$ 
corresponds
to a constant value of the quantity $\ell(\ell+1)C_\ell$. For
this reason it is usually in terms of this combination
of $\ell$ and $C_\ell$ that the data from the CMBR are 
represented.

The WMAP team \cite{wmap3} has found that the two point correlation
function $C(\theta)$, simply obtained from the $C_\ell$, nearly
vanishes on scales greater than about 60 degrees, contrary to what the
standard theories predict, and in agreement with the same finding
obtained from COBE data about a decade earlier \cite{cobe}. Recently
it was confirmed \cite{shwarz2}, by considering the WMAP three- and
five-year maps, the lack of correlations on angular scales greater
than about 60 degrees at a level that would occur only in 0.025 per
cent of realizations of the LCDM model.  Moreover, particularly
puzzling are the alignments of low multi-poles with the solar system
features \cite{shwarz1,huterer}, i.e. the alignments between the
quadrupole and octopole and between these multipoles and the geometry
of the Solar System. This would imply that CMBR anisotropy should be
correlated with our local habitat. A possible conclusion is that the
observed correlations seem to hint that there is contamination by a
foreground or that there is an important systematic effect in the data
\cite{cover2}. More recently Cover \cite{cover1} found that there are
substantial differences at large scale (low-$\ell$) between the WMAP
and the preliminary maps provided by the Planck satellite, concluding
that the presence of systematic effects at large angular separation
could possibly explain the peculiar features found in the WMAP and
COBE data (see also \cite{cover2} and references therein).  It was
then found that the amplitudes of the low multipoles measured from the
preliminary Planck satellite data, are significantly lower than that
reported by the WMAP team \cite{ll2}.  Actually it was concluded that
the Planck first light survey image strongly supports the artificial
origin of quadrupole observed in WMAP maps and that the real CMBR
quadrupole is most possibly near zero.

In summary from the observational point of view, at present one is not
able to determine whether fluctuations in the radiation and matter
density fields really show the crucial super-homogeneous
features. However if it will be confirmed by the Planck mission that
the temperature PS $C_\ell$ of the CMBR does not decay as $1/(\ell
(\ell+1))$ at low $\ell$, this would put in troubles the whole
scenario of galaxy formation models based on the inflationary
paradigm, i.e. the ``scale-invariant'' nature of matter density
fluctuations.

\section{Conclusions} 

In summary galaxy structures are highly inhomogeneous up to scales of
100 Mpc/h, or more as indicated by field-to-field fluctuations. This
situation is in contradiction with the prediction of the LCDM model in
which the scale beyond which the distribution should become uniform is
about 10 Mpc/h.  We have discussed the problems related to finite size
effects which must be carefully considered in the analysis of spatial
correlations. These finite-size effects are responsible for the
contradictory results obtained by different authors with different
statistical methods or by considering different galaxy samples.

In addition we discussed that the super-homogeneous nature of matter
and radiation has not been detected, neither in galaxy catalogs nor in
the CMBR anisotropies. In the latter case the situation is very
puzzling as indicated by recent results. This situation calls for a
more deep analysis of the foundations of the standard model of galaxy
formation.

\section*{Acknowledgments}
I warmly thank the organizers of the workshop on ``Cosmic Structure
and Evolution''. I acknowledge stimulating discussions with Keith
Cover, Gary Hinshaw, Dominik J. Schwarz and Glenn Starkman. I also
thank my collaborators with whom different parts of the work here
presented have been developed: Tibor Antal, Yuri V. Baryshev, Andrea
Gabrielli, Michael Joyce, Martin L\'opez-Corredoira and Nickolay
L. Vasilyev.

\end{document}